\documentclass[10pt,twocolumn,letterpaper]{article}

\usepackage{wacv}
\usepackage{times}
\usepackage{epsfig}
\usepackage{graphicx}
\usepackage{amsmath}
\usepackage{amssymb}
\usepackage{floatrow}

\newcommand{\saeed}[1]{{\color{black} #1}}

\def\wacvPaperID{960} 

\wacvfinalcopy 

\ifwacvfinal
\def\assignedStartPage{1} 
\fi

\ifwacvfinal
\usepackage[breaklinks=true,bookmarks=false]{hyperref}
\else
\usepackage[pagebackref=true,breaklinks=true,colorlinks,bookmarks=false]{hyperref}
\fi

\ifwacvfinal
\else
\fi

\begin{document}


\title{Benefiting from Bicubically Down-Sampled Images for Learning \\ Real-World Image Super-Resolution}

\author{Mohammad Saeed Rad$^{1*}$ \quad \quad

Thomas Yu$^{1}\thanks{Equal contributions.}$ \quad \quad

Claudiu Musat$^{2}$ \quad \quad

Haz{\i}m Kemal Ekenel$^{1,\, 3}$\\

Behzad Bozorgtabar$^{1}$ \quad \quad

Jean-Philippe Thiran$^{1}$
\and
$^{1}$LTS5, EPFL, Switzerland \quad \quad
$^{2}$Digital Lab, Swisscom, Switzerland \quad \quad
$^{3}$SiMiT Lab, ITU, Turkey\\
{\tt\small \{saeed.rad, firstname.lastname\}@epfl.ch \qquad \{firstname.lastname\}@swisscom.com}
}

\maketitle


\begin{abstract}
\saeed{Super-resolution (SR) has traditionally been based on pairs of high-resolution images (HR) and their low-resolution (LR) counterparts obtained artificially with bicubic downsampling. However, in real-world SR, there is a large variety of realistic image degradations and analytically modeling these realistic degradations can prove quite difficult. In this work, we propose to handle real-world SR by splitting this ill-posed problem into two comparatively more well-posed steps. First, we train a network to transform real LR images to the space of bicubically downsampled images in a supervised manner, by using both real LR/HR pairs and synthetic pairs. Second, we take a generic SR network trained on bicubically downsampled images to super-resolve the transformed LR image. The first step of the pipeline addresses the problem by registering the large variety of degraded images to a common, well understood space of images. The second step then leverages the already impressive performance of SR on bicubically downsampled images, sidestepping the issues of end-to-end training on datasets with many different image degradations. We demonstrate the effectiveness of our proposed method by comparing it to recent methods in real-world SR and show that our proposed approach outperforms the state-of-the-art works in terms of both qualitative and quantitative results, as well as results of an extensive user study conducted on several real image datasets.}
\end{abstract}

\begin{figure}[ht]
\begin{center}
   \includegraphics[width=0.9\linewidth]{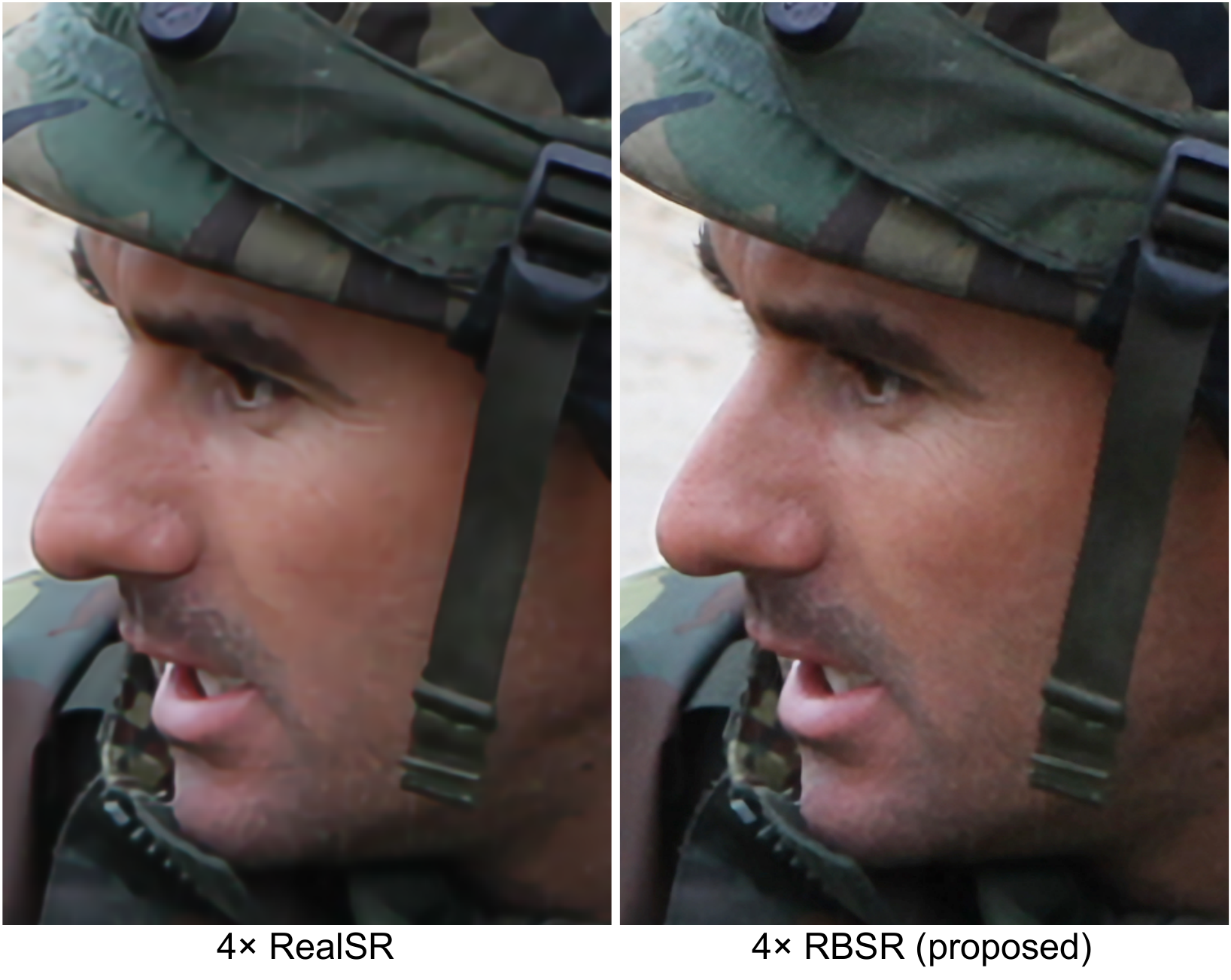}
\end{center}
   \caption{An example SR produced by our system on a real-world LR image, for which \textbf{no higher resolution/ground-truth is available}. Our method is compared against the RealSR \cite{cai2019toward} method, a state-of-the-art of real SR method trained in a supervised way on real low-resolution and high-resolution pairs. The low-resolution image is taken from HR images in the DIV2K validation set~\cite{divk2_ref}.}
   \vspace{-2mm}
\label{fig:cover_pic}
\end{figure}

\begin{figure*}[ht]
\begin{center}
   \includegraphics[width=0.95\linewidth]{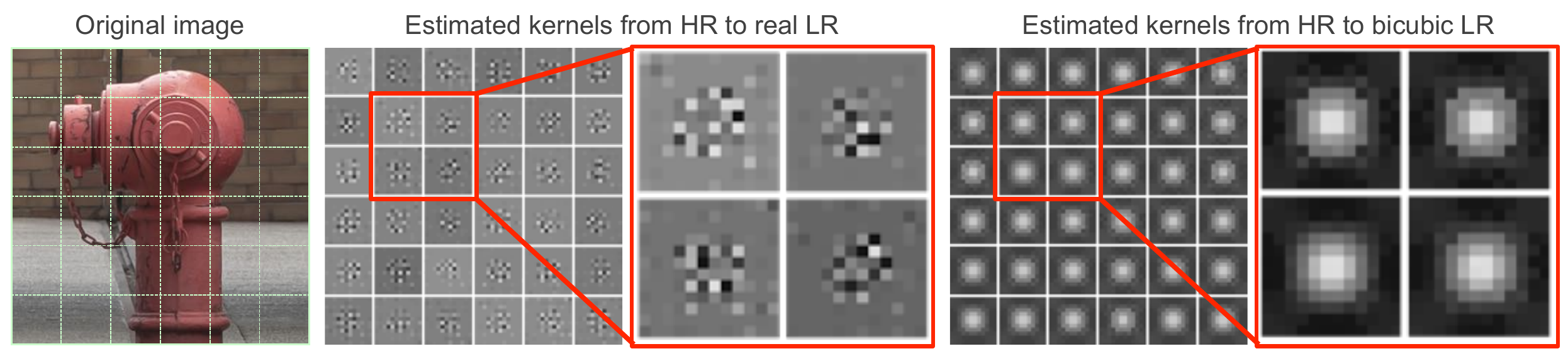}
\end{center}
   \caption{Downsampling kernels estimated patchwise on a RealSR~\cite{cai2019toward}  LR image and the same image bicubically downsampled from the HR image. Estimations were done using least squares optimization with regularization on the kernel using the LR and HR images, assuming the standard degradation model of kernel convolution followed by subsampling. We can see that the RealSR LR images are difficult to estimate with the standard image degradation model.}
\label{fig:visualization_kernles}
\vspace{-3mm}
\end{figure*}

\section{Introduction}
\saeed{Super resolution is the generally, ill-posed problem of reconstructing high-resolution (HR) images from their low-resolution (LR) counterparts. Generally SR methods restrict themselves to super-resolving LR images downsampled by a simple and uniform degradations (i.e, bicubic downsampling) \cite{srsurvey,wang2019deep,anwar2019deep,dong2014learning,tai2017memnet}. Although the performance of these methods on artificially downsampled images are quite impressive \cite{wang2018esrgan,dai2019second}, applying these methods on real-world SR images, with unknown degradations from cameras, cell-phones, etc. often leads to poor results \cite{cai2019toward,lugmayr2019unsupervised}. The real-world SR problem is then to super-resolve LR images downsampled by unknown, realistic image degradations \cite{lugmayr2019aim}. 

Recent works try to resemble realistic degradations by acquisition instead of artificial downsampling, such as hardware binning, where LR corresponds to a coarser grid of photoreceptors~\cite{DBLP}, or camera focal length changes, which changes the apparent size of an object in frame~\cite{cai2019toward}. These approaches could propose very limited number of physically real low and high-resolution pairs and their degradation models are limited to very few acquisition hardwares.

As shown in \cite{efrat2013accurate}, correct modeling of the image degradation is crucial for accurate super-resolution. A general, analytical model for image degradation which is commonly assumed is $\mathbf{y}=(\mathbf{x}*\mathbf{k})\downarrow_s + N$, 
where $\mathbf{y}$ is the LR image, $\mathbf{x}$ is the HR image, $*$ denotes convolution, $\mathbf{k}$ is the blur kernel, $N$ is noise, and $\downarrow_s$ denotes downsampling by a factor $s$. However, as can be seen in Figure \ref{fig:visualization_kernles}, these convolutional models are only approximations to the true, real degradations.

Recently, there has been a push to account for more realistic image degradations through physical generation of datasets with real LR to HR pairs \cite{cai2019toward}, synthetically generating real LR to HR pairs through unsupervised learning or blind kernel estimation \cite{lugmayr2019unsupervised,zhou2019kernel},
and simulating more complex image degradation models such as in equation 1, with and without restrictions on $\mathbf{k}$ and $\downarrow_s$ \cite{gu2019blind,zhang2019deep}. The pipelines of these approaches generally have the ultimate goal of training an end-to-end network to take as input a ``real" image and output a SR image. Although these approaches result in better reconstruction quality, the real challenge of the real-world LR to HR problem is not only limited to a lack of real LR and HR pairs; the large variety of degraded images and the difficulty in accurately modeling the degradations makes realistic SR even  more ill-posed than SR based on bicubically down-sampled images~\cite{zhang2018learning}.\\

\noindent
\textbf{Main idea} We propose to address real world SR with a two-step approach, which we call Real Bicubic Super-Resolution (RBSR). RBSR generally decomposes the difficult problem of real world SR into two, sequential subproblems: \textbf{1-} Transformation of the wide variety of real LR images to a single, tractable LR space. \textbf{2-} Use of generic, bicubic SR networks with the transformed LR image as input.

We choose to transform real LR images to the common space of bicubically downsampled images because of two main advantages. First, bicubic images are tractably generated with the standard convolutional model of image degradation, therefore the inverse transform is less ill-posed comparing to the cases of arbitrary/unknown degradations. Second, we can leverage the already impressive performance of SR networks trained on bicubically downsampled images, thanks to the availability of huge SR image datasets using bicubic kernels (see Figure \ref{fig:cover_pic}).

In summary, our contributions are as follows:
\begin{enumerate}
    \item  We use a GAN to train a CNN-based image-to-image translation network, which we call a ``bicubic look-alike generator'', to map the distribution of real LR images to the easily modeled and well understood distribution of bicubically downsampled LR images. We use a SR network with the transformed LR image by our proposed bicubic look-alike generator as input to solve the \textbf{real-world super-resolution} problem.
    \item To this end, and for the consistency of the bicubic look-alike generator, we propose a novel copying mechanism, where the network is fed with identical, bicubically downsampled images as both input and ground-truth during training; this way, the network loses its tendency to merely sharpen the input images, as realistic low-resolution images usually seem to be much smoother. 
    \item We train our bicubic look-alike generator by using an extended version of perceptual loss, where its feature extractor is specifically trained for SR task and on bicubically downsampled images. The proposed ``bicubic perceptual loss'' is shown to have less artifacts.
    \item We demonstrate the effectiveness of the proposed two-step approach by comparing it to an end-to-end setup, trained in the same setting. Furthermore, we show that our proposed approach outperforms the state-of-the-art works in terms of both qualitative and quantitative results, as well as results of an extensive user study conducted on several real image datasets.
\end{enumerate}

In essence, training models on paired datasets of real LR and HR pairs requires expensive collection of big datasets; in addition, training a single model on multiple degradations for SR is ill-posed/vulnerable to instability \cite{zhang2018learning}. Training on synthetic datasets coming from analytical degradation models have the benefit of much larger datasets and an easier task for the network, at the cost of being less realistic. However, this approach still has the ill-posedness problem of training on multiple degradations. In RBSR, we try to simultaneously keep the added information from realistic LR images and the impressive performance of SR networks on single, well-defined degradations. 

}


\section{Related Work}
The vast majority of prior work for Single image super-resolution (SISR) focuses on super-resolving low-resolution images which are artificially generated by bicubic or Gaussian downsampling as the degradation model. We consider that recent research on addressing real-world conditions can be broadly categorized into two groups. The first group proposes to physically generate new, real LR and HR pairs and/or learn from real LR images in supervised and unsupervised ways (Section \ref{subsec:real-world-sr-real}). The second group extends the standard bicubic downsampling model, usually by more complex blur kernels, and generates new, synthetic LR and HR pairs (Section \ref{subsec:real-world-sr-extended}).

\subsection{Real-World SR through real data}
\label{subsec:real-world-sr-real}
Some recent works \cite{chen2019camera,cai2019toward} propose to capture real LR/HR image pairs to train SR models under realistic settings. However, the amount of such data is limited. The authors in \cite{cai2019toward,chen2019camera} proposed to generate real, low-resolution images by taking two pictures of the same scene, with camera parameters all kept the same, except for a changing camera focal length. Hence, the image degradation corresponds to "zooming" out of a scene. They generate a dataset of real LR and HR pairs according to this procedure and show that bicubically trained SR models perform poorly on super-resolving their dataset. Since this model's image degradation can be modeled as convolution with a spatially varying kernel, they propose to use a kernel prediction network to super-resolve images. In \cite{lugmayr2019unsupervised}, the authors perform unsupervised learning to train a generative adversarial network (GAN) to map bicubically downsampled images to the space of real LR images with two unpaired datasets of bicubically downsampled images and real LR images. They then train a second, supervised network to super-resolve real LR images, using the transformed bicubically downsampled images as the training data. In a similar work, \cite{bulat2018learn} trains a GAN on face datasets, for the specific face SR task, but their approach relies on unrealistic blur-kernels.

In \cite{bellkligler2019blind}, the authors model image degradation as convolution over the whole image with a single kernel, followed by downsampling. Given a LR image, they propose a method to estimate the kernel used to downsample the image solely from subpatches of the image by leveraging the self-similarity present in natural images. This is done by training a GAN, where the generator produces the kernel and the discriminator is trained to distinguish between crops of the original image and crops which are downsampled from original image using this estimated kernel. This method relies on the accuracy of the standard convolutional model of downsampling, which is shown to not hold for RealSR images in Figure \ref{fig:visualization_kernles}. Further, the estimation of the kernel and subsequent SR are quite time consuming in comparison to supervised learning based methods; the calculation of the kernel alone for a $1000\times 1000$ image can take more than three minutes on a GTX 1080 TI. In addition, their method constrains the size of the input images to be "large enough" since they need to downsample the input images during training. In \cite{yuan2018unsupervised}, the authors propose an unsupervised cycle-in-cycle GAN, where they create one module for converting real LR images to denoised, deblurred LR images and one module for SR using these Clean LR images. They then tune these networks simultaneously in an end-to-end fashion, which causes this intermediate representation of the LR image to deviate from their initial objective.

\begin{figure*}[h]
\vspace{-5mm}
\begin{center}
   \includegraphics[width=0.75\linewidth]{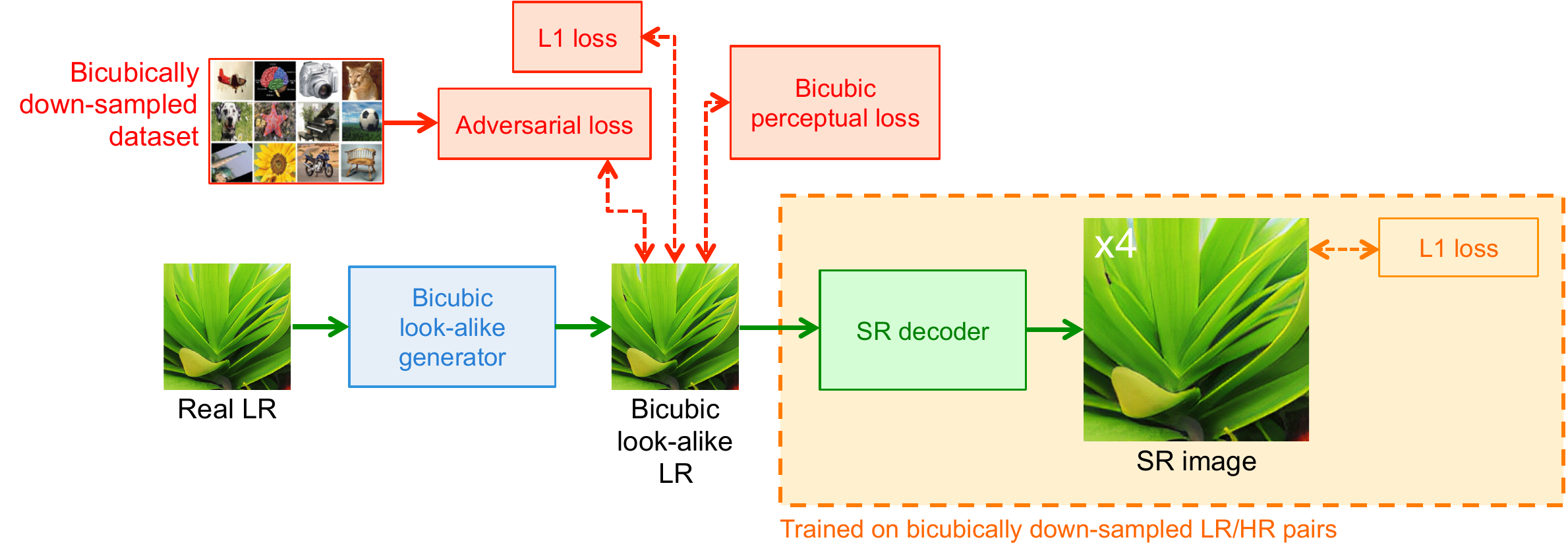}
\end{center}
   \caption{We propose a two-step pipeline for real world SR. First, we transform real LR images to bicubically downsampled looking images through our bicubic look-alike generator. We then pass the transformed image as input to a generic SR decoder trained on bicubically downsampled images.}
\label{fig:pipeline}
\vspace{-5mm}
\end{figure*}

\subsection{Real World SR through extended models}
\label{subsec:real-world-sr-extended}
In \cite{zhang2019deep}, the authors extend the bicubic degradation model by modeling image degradation as a convolution with an arbitrary blur kernel, followed by bicubic downsampling. They embed the super-resolution in an alternating iterative scheme where analytical deblurring is alternated with applying a SR network trained on bicubically downsampled images. Although this method generalizes to arbitrary kernels, one has to provide the kernel and the number of iterations as an input to the pipeline.
In \cite{gu2019blind}, the authors extend the bicubic degradation model by modeling image degradation as a convolution with a Gaussian blur kernel, followed by bicubic downsampling. They use an iterative scheme using only neural networks, where at each iteration the pipeline produces both the SR image and an estimate of the corresponding downsampling kernel. 
In \cite{zhou2019kernel}, the authors also model image degradations as convolution with a blur kernel followed by bicubic downsampling. They estimate the blur kernel using a pre-existing blind deblurring method on a set of "real" images which are bicubically upsampled; they use the same dataset of low quality cell-phone pictures used in \cite{lugmayr2019unsupervised}. They then train a GAN to generate new, realistic blur kernels using the blindly estimated blur kernels. Finally, they generate a large synthetic dataset using these kernels and train an end-to-end network on this dataset to perform SR. These three methods all rely on an analytical model for image degradation as well as being reliant on restrictive kernels or blind kernel estimation.


\section{Methodology}
\subsection{Overall pipeline}

RBSR consists of two steps; first, we use a Convolutional Neural Network (CNN)-based network, namely the bicubic look-alike image generator, whose objective is to take as input the real LR image and transform it into an image of the same size and content, but which looks as if it had been downsampled bicubically rather than with a realistic degradation. We call this output the bicubic look-alike image. Second, we use any generic SR network trained on bicubically downsampled data to take as input the transformed LR image and output the SR image. Figure \ref{fig:pipeline} shows an overview of our proposed pipeline. We restrict the upsampling factor to four. In the following subsections, we describe each component of our pipeline in more details. 

\subsection{Bicubic look-alike image generator}
The bicubic look-alike image generator is a CNN, trained in a supervised manner. The main objective of this network is to transform real LR images to bicubic look-alike images. In this section, we present its architecture in detail. Then, we introduce a novel perceptual loss used to train it. Finally, we also introduce a novel copying mechanism used during training to make this transformation consistent. 

\subsubsection{Architecture}
\begin{figure*}[t]
\vspace{-4mm}
\begin{center}
   \includegraphics[width=0.75\linewidth]{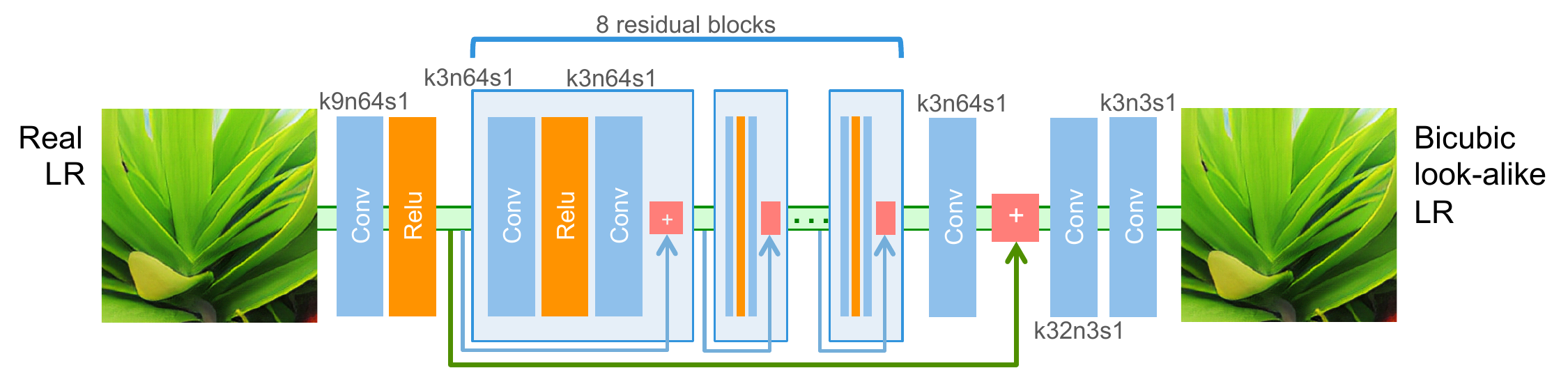}
\end{center}
   \caption{Schematic diagram of the bicubic-alike decoder. We train the decoder using our new bicubic perceptual loss, alongside standard $L_1$ and adversarial losses. In this schema, $k$, $n$ and $s$ correspond to kernel size, number of feature maps and stride size, respectively.}
\label{fig:archi}
\vspace{-1mm}
\end{figure*}

The architecture of the bicubic look-alike generator is shown in Figure \ref{fig:archi}. The generator is a feed-forward CNN, consisting of convolutional layers and several residual blocks, which has shown great capability in image-to-image translation tasks \cite{liu2017unsupervised}. The real low-resolution image $I^{Real-LR}$ is passed through the first convolutional layer with a ReLU activation function with a 64 channel output. This output is subsequently passed through 8 residual blocks. Each block has two convolutional layers with $3\times3$ filters and $64$ channel feature maps. Each one is followed by a ReLU activation. By using a long skip connection, the output of the final residual block is concatenated with the features of the first convolutional layer. Finally, the result is filtered by a last convolution layer to get the the 3-channel bicubic look-alike image ($I^{Bicubic-LR}$).

\subsubsection{Loss functions}
In the bicubic look-alike generator, we use a loss function ($\mathcal{L}_{total}$) composed of three terms: 1- Pixel-wise loss ($\mathcal{L}_{pix.wise}$), 2- adversarial loss, and 3- our novel bicubic perceptual loss function ($\mathcal{L}_{bic.perc.}$). The overall loss function is given by: 
\vspace{-1mm}
\begin{equation}
\mathcal{L}_{total} = \alpha \mathcal{L}_{pix.wise} + \beta \mathcal{L}_{bic.perc.} + \gamma \mathcal{L}_{adv} \quad,
\label{eq:cf2}
\end{equation}

\noindent
where $\alpha$, $\beta$ and $\gamma$ are the corresponding weights of each loss term used to train our network. In the following, we present each term in detail:

\textbf{$\bullet$ Pixel-wise loss.} We use the $L_1$ norm of the difference between predicted and ground-truth images as this has been shown to improve results compared to the $L_2$ loss \cite{zhao2015loss}.

\textbf{$\bullet$ Adversarial loss.} This loss measures how well the image generator can fool a separate discriminator network, which originally was proposed to reconstruct more realistic looking images for different image generation tasks~\cite{goodfellow2014generative,paper_twitter_0,8756632,BOZORGTABAR20191}. However, in this work, as we are feeding the discriminator with bicubically downsampled images as the ``real data'', it results in images which are indistinguishable from bicubically downsampled images.
The discriminator network used to calculate the adversarial loss is similar to the one presented in~\cite{paper_twitter_0}; it consists of a series of convolutional layers with the number of channels of the feature maps of each successive layer increasing by a factor of two from that of the previous layer, up to 512 feature maps. The result is then passed through two dense layers, and finally, by a sigmoid activation function. The discriminator classifies the images as either ``bicubically downsampled image'' (real) or ``generated image''(fake).

\textbf{$\bullet$ Bicubic perceptual loss.}
Perceptual loss functions \cite{johnson2016perceptual,ledig2017photo} tackle the problem of blurred textures caused by optimization of using per-pixel loss functions and generally result in more photo-realistic reconstructions. In this work, we take inspiration from this idea of perceptual similarity by introducing a novel perceptual loss. 

However, instead of using a pre-trained classification network, e.g. VGG~\cite{paper_vgg} for the high-level feature representation, we use a pre-trained SR network trained on bicubically down-sampled LR/HR pairs. In particular, we use the output of the last residual block of our SR network, presented in Section~\ref{sec:SR_generator}, to map both HR and SR images into a feature space and calculate their distances. The bicubic perceptual loss term is formulated as:

\begin{align}
\begin{split}
\mathcal L_{bic.\_perc.} = \frac{1}{W_{i,j}H_{i,j}}\sum_{x=1}^{W_{i,j}}\sum_{y=1}^{H_{i,j}}\Big( \phi_{k}^{SR}\left ( I^{Bicubic-LR} \right ) \\
 - \phi_{k}^{SR}\left ( I^{T-LR} \right ) \Big)^{2}, \qquad \qquad \qquad
\label{eq:1}
\end{split}
\end{align}

\noindent
where $W_{i,j}$ and $H_{i,j}$ denote the dimensions of the respective feature maps. $\phi_{k}^{SR}$ indicates the output feature map of the $k$-th residual block from the SR decoder and $I^{T-LR}$ denotes the transformed LR image. We conjecture that using a SR feature extractor, which is specifically trained for SR task and on bicubically down-sampled images, will better reflect features corresponding to the characteristics of bicubically downsampled images than using a feature extractor trained for image classification. 

\begin{figure}[b]
\vspace{-4mm}
\begin{center}
   \includegraphics[width=0.7\linewidth]{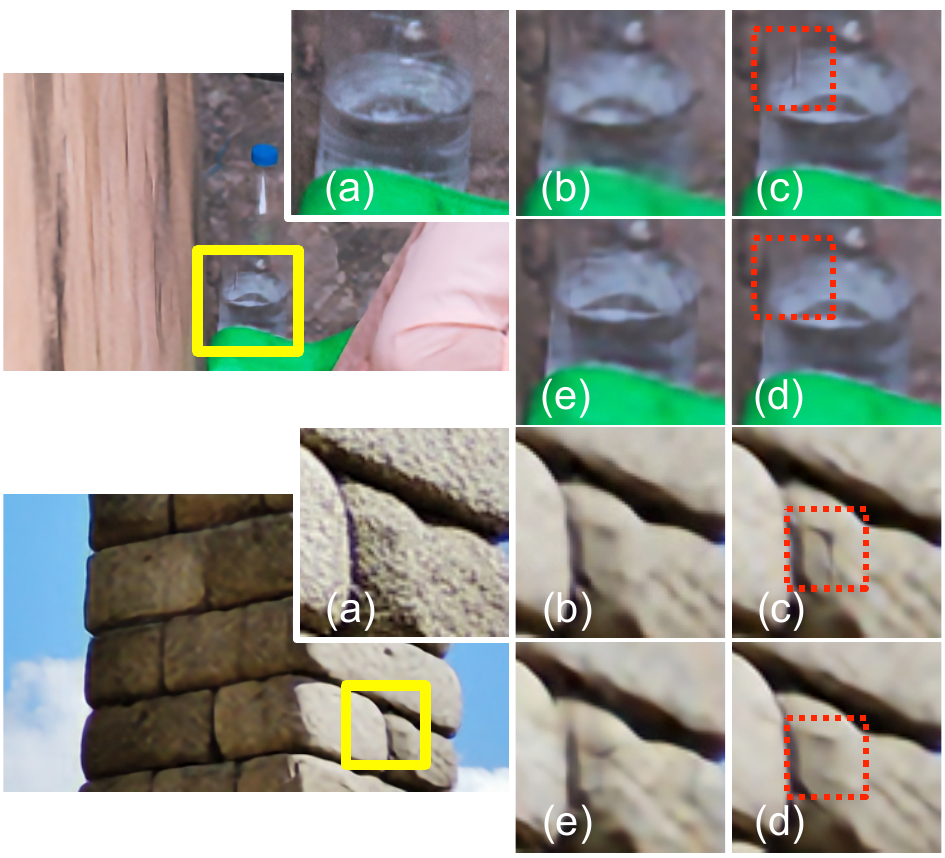}
\end{center}
   \caption{The effectiveness of using bicubic perceptual loss: (a) HR image, (b) Only L1 loss, (c) perceptual loss, (d) bicubic perceptual loss, and (e) bicubic perceptual loss + adversarial loss. Red boxes show how using bicubic perceptual loss (c) decreases artifacts comparing to using conventional perceptual losses (d), while still producing sharper edges comparing to only using $L_1$ loss.}
\label{fig:bic_perc_loss}
\vspace{-4mm}
\end{figure}

In Figure ~\ref{fig:bic_perc_loss}, we compare the effect of using the standard perceptual loss which uses a pre-trained classification network versus our bicubic perceptual loss. Note that the standard perceptual loss introduces artifacts in the transformed LR image which are avoided by the bicubic perceptual loss. Further, we see that using the bicubic perceptual loss produces sharper edges as compared to using just the $L_1$ loss.

\subsubsection{Copying mechanism}
\label{subsec:copyingmechanism}
Bicubically downsampled images are in general seem to be much sharper than realistic low-resolution images, therefore, training network by real LR images gives it this tendency to merely sharpen the input images instead of learning bicubic characteristics. To address this issue, we want the network to be consistent and apply minimal sharpening to already sharp images. To this end, we utilize a novel copying mechanism, where the network is periodically fed with identical, bicubically downsampled images as both input and output during training. This is done in order to prevent the network from just learning to sharpen images, as this can cause oversharpening or amplification of artifacts.

\begin{figure}[ht]
\vspace{-3mm}
\begin{center}
   \includegraphics[width=0.7\linewidth]{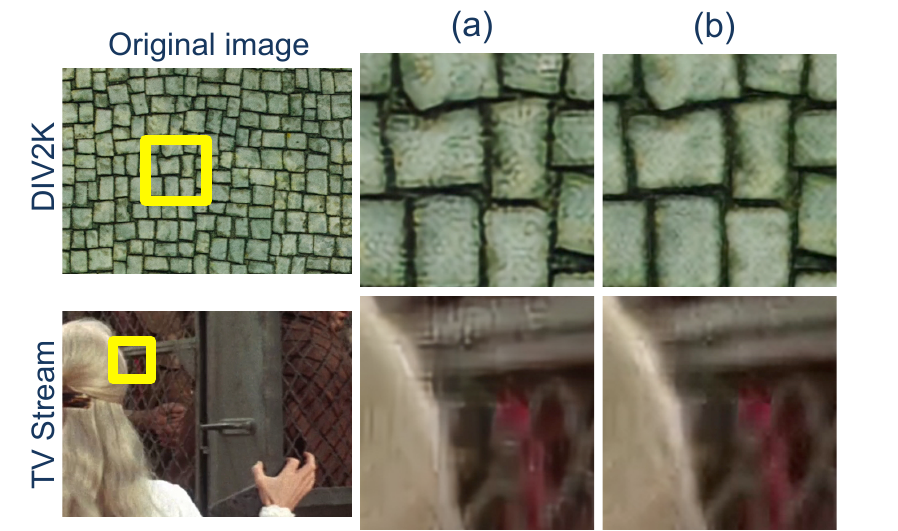}
\end{center}
   \caption{Example images generated without (a) and with (b) the copying mechanism during training. We can clearly see that without the copying mechanism, resulting images suffer from oversharpening and artifact amplification.}
\label{fig:copying_effect}
\vspace{-3mm}
\end{figure}

In Figure \ref{fig:copying_effect} we compare the outputs of the network trained with and without the copying mechanism. We can see clearly that training without the copying mechanism results in severe over-sharpening of the output image. 

\subsection{SR generator}
The second step of our pipeline is to feed the output of our bicubic-like image generator as the input to any SR network trained on bicubically downsampled images. For simplicity, we use a network based on EDSR~\cite{paper_edsr}. The EDSR architecture is composed of a series of residual blocks bookended by convolutional layers. Crucially, batch normalization layers are removed from these blocks for computational efficiency and artifact reduction. For simplicity, as well as decreasing training/inference time, we only use 16 residual blocks, as compared to the 32 residual blocks used in EDSR. This generator is trained on DIV2K training images (track 1: bicubically downsampled images and HR pairs) and by using the $L_1$ loss function. We refer the reader to the supplementary material for more details about the network architecture. 
\label{sec:SR_generator}

\subsection{Training parameters}
\label{sec:training_params}
\noindent
\textbf{Bicubic look-alike generator}
For the training data, as input, we use 400 RealSR \cite{cai2019toward} and 400 DIV2K Track 2~\cite{divk2_ref} LR images. The RealSR dataset contains real LR-HR pairs, captured by adjusting the focal length of a camera and taking pictures from the same scene. Track 2 images are downsampled using unknown kernels. As the desired output is the bicubic look-alike image, we use the bicubically downsampled RealSR and the bicubically downsampled DIV2K (track 1) images as the ground truth for the training inputs. In addition, as described in Section~\ref{subsec:copyingmechanism}, we add 400 bicubically downsampled images from DIV2K, identical for both input and ground-truth, to make the generator consistent and avoid oversharpening or artifact amplification. We use the same 400 bicubically downsampled images from DIV2K as the real input of the discriminator. At each epoch, we randomly cropped the training images into $128\times 128$ patches. The mini-batch size in all the experiments was set to $16$.
The training was done in two steps; first, the SR decoder was pre-trained for 1000 epochs with only the $L_1$ pixel-wise loss function. Then the proposed bicubic perceptual loss function, as well as the adversarial loss, were added and the training continued for 3000 more epochs. The weights of the $L_1$ loss, bicubic perceptual loss and adversarial loss function ($\alpha$, $\beta$ and $\gamma$) were set to $1.0$, $3.0$, and $1.0$ respectively. The Adam optimizer \cite{kingma2014adam} was used during both steps. The learning rate was set to $1 \times 10^{-4}$ and then decayed by a factor of 10 every 800 epochs. We also alternately optimized the discriminator with similar parameters to those proposed by~\cite{paper_twitter_0}.

\noindent
\textbf{SR generator}
The SR decoder is also trained in a single step for 4000 epochs and using the $L_1$ loss function. For the training data, we only use track 1 images of DIV2K, which consists of 800 pairs of bicubically downsampled LR and HR images. Similar to the training of the bicubic look-alike generator, the Adam optimizer was used for the optimization process. The learning rate was set to $1 \times 10^{-3}$ and then decayed by a factor of 10 every 1000 epochs.

\noindent
\textbf{End-to-end baseline}
To investigate the effectiveness of RBSR, which super-resolves a given input in two steps, we also fine-tune the EDSR architecture with the same datasets used to train the bicubic look-alike generator. This dataset consists of 400 RealSR and 400 DIV2K Track 2 LR and HR pairs. We further noticed that the inclusion of 400 bicubically downsampled LR and HR pairs in this dataset adds more robustness to the performance. In order to keep the same number of parameters as in the RBSR pipeline, we increase the number of residual blocks of this end-to-end generator to 24. The training parameters used for this baseline is similar to the ones used in~\cite{paper_edsr}.

\begin{figure*}
\vspace{-6mm}
\begin{center}
   \includegraphics[width=0.9\linewidth]{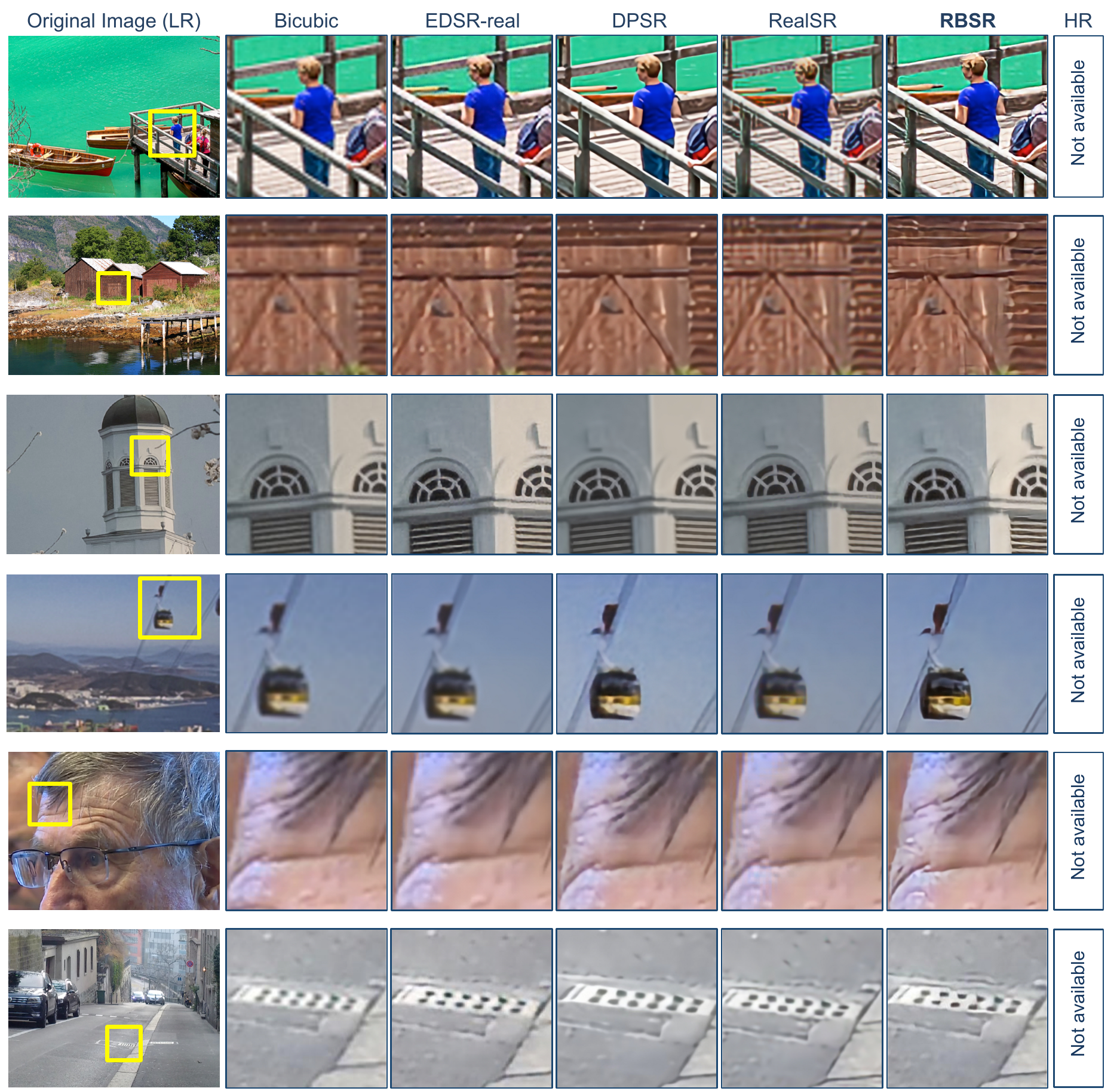}
\end{center}
   \caption{
   Qualitative results of $\times 4$ SR on a subset of the DIV2k~\cite{divk2_ref} (Rows 1-2), RealSR HR~\cite{cai2019toward} (Rows 3-4), TV Streams (Row 5), and DPED cell-phone images~\cite{dped} (Row 6). Results from left to right: bicubic, EDSR~\cite{paper_edsr} fine-tuned with real LR and HR pairs, DPSR~\cite{zhang2019deep}, RealSR~\cite{cai2019toward}, and RBSR (ours). Please note that no ground-truth is available for these images. More results can be seen in the supplementary material. \textbf{Zoom in for the best view.}\vspace{-1mm}}
\label{fig:qualitative_results}
\vspace{-2mm}
\end{figure*}

\vspace{-3mm}
\section{Experimental results}
\vspace{-1mm}
In this section, we compare RBSR to several SOTA algorithms (CVPR 2019, ICCV 2019) in real-world SR both qualitatively and quantitatively. We show standard distortion metrics for the datasets with ground truth, and we show a comprehensive user study conducted over six image datasets with varying image quality and degradations. In all cases, we use an upsampling factor of four.

We emphasize that the distortion metrics are not directly correlated to the perceptual quality as judged by human raters~\cite{pirm_challenge,paper_twitter_0,paper_enhanced,Rad_2019_ICCV,paper_segmentation,RAD2020304}; the super-resolved images could have higher errors in terms of the PSNR and SSIM metrics, but still generate more appealing images. Moreover, the RealSR images represent only a limited group of realistic images from Nikon and Canon cameras. Therefore, we validate the effectiveness of our approach by qualitative comparisons and by an extensive user study in the following sections.

\saeed{\subsection{Test images}
\subsubsection{Lack of ground-truth in real-world SR}
\label{sec:im_groundtruth}
One of the main challenges of real-world SR is the lack of real low and high resolutions pairs, for both training and testing. As mentioned previously, most of the known benchmarks in super-resolution had no choice but using a known kernel to create a counterpart with lower resolution. To the best of our knowledge RealSR~\cite{cai2019toward} is the only dataset with real images of the same scenes with different resolutions: their LR and HR images are generated by taking two camera pictures of the same scene, but changing the focal length of the camera between the two pictures. Hence, both are real images, but with the RealSR LR being degraded with the degradation from changing the focal length of the camera (zooming out). DIV2K Unknown kernel LR images \cite{divk2_ref} is another attempt to create pairs of real low and high-resolutions images. They generate synthetically real low and high resolution images by using unknown/random degradation operators.

\begin{figure*}[t]
\vspace{-6mm}
\begin{center}
   \includegraphics[width=1.0\linewidth]{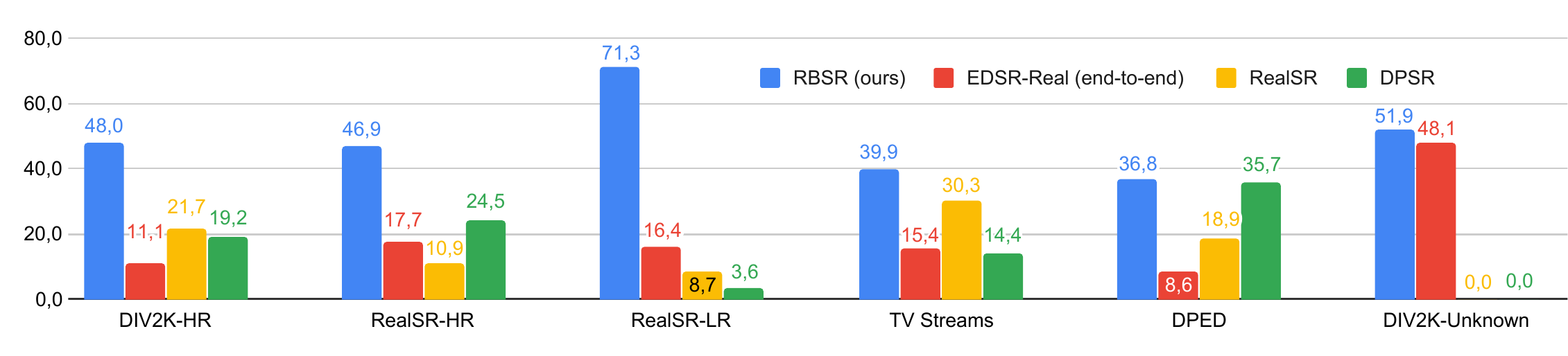}
\end{center}
   \caption{Results of the user study comprising forty one people, comparing EDSR~\cite{paper_edsr}, fine-tuned with real LR and HR pairs, DPSR~\cite{zhang2019deep}, RealSR~\cite{cai2019toward}, and RBSR (ours), on six different datasets: DIV2K HR \cite{divk2_ref}, RealSR \cite{cai2019toward} HR, RealSR LR, TV Stream images, DPED \cite{dped} Mobile Phone images, and DIV2K Unknown Kernel LR.\vspace{-3mm}}
\label{fig:user_study}
\vspace{-4mm}
\end{figure*}

\subsubsection{Images without ground-truth}
In addition to RealSR LR and DIV2K Unknown kernel datasets, we also evaluate our method on four datasets of real images, without having any ground-truth as it is the main focus of real-world SR task: 1- RealSR \cite{cai2019toward} HR test images, 2- DIV2K HR \cite{divk2_ref} validation images (real), 3- DPED \cite{dped} Mobile Phone images, 4- TV Stream images (unknown, depending on the original content of the TV).} The DPED Mobile Phone dataset is a dataset of real images where cell-phones were used to take pictures of same scenes. The TV stream images are decoded images from an actual TV channel stream at HD $(1920\times 1080)$ resolution; our acquisition algorithm captured one image every ten minutes over a period of two days, to ensure that our these test images cover different types of content. We note that no information is available about their type of degradations, as the original resolutions of the contents before streaming are unknown. Further, we note that we only have the ground-truth high-resolution images for the DIV2K Unknown Kernels images and the RealSR LR images.

\subsection{Quantitative results}
In this work, calculating distortion metrics such as PSNR and SSIM is not possible for test images that truly reflect the real-world problem (original images from smartphones, TV streams, etc.), as in real cases the downsampling operator is not known and therefore no ground-truth is available. RealSR~\cite{cai2019toward} is the only dataset with physically produced high and low-resolution image pairs. Readers can refer to our \textbf{supplementary material A} to find PSNR, SSIM and perception index (PI) metric evaluated by using this dataset.

\subsection{Qualitative comparison}
\label{sec:qual_com}
For the qualitative comparison, we compare the following real world SR algorithms: 1- RBSR (Ours), 2- EDSR-real: the EDSR~\cite{paper_edsr} network trained end-to-end on the same data/settings as RBSR, 3- The pretrained RealSR network~\cite{cai2019toward}, and 4- The pre-trained DPSR network with default settings for real-world SR~\cite{zhang2019deep}. We compare with the end-to-end EDSR network in order to show the efficacy of splitting the problem into two steps. We compare to RealSR and DPSR as they are two of the most recent state-of-the-art algorithms. We use their pre-trained models along with the default settings for real images they provide\footnote{https://github.com/csjcai/RealSR}$^{,}$\footnote{https://github.com/cszn/DPSR}. In Figure ~\ref{fig:qualitative_results}, we show qualitative results on a random subset of the image datasets described in the previous sections.

\vspace{-1mm}
\subsection{User study}
We also conducted a user study comprising forty one people in order to gauge the perceptual image quality of SR images using the image datasets described in the previous section. We chose five images randomly from each dataset, with thirty total images. For each image, the users were shown four SR versions of the image, each corresponding to the real-world SR algorithms being compared. Users were asked to select which SR image felt more realistic and appealing. The images were shown to users in a randomized manner. As the datasets reflect a wide range of image quality, etc., we show the evaluations of the algorithms for each dataset separately. Our metric of evaluation for the algorithms is the percent of votes won. We show the results of the user study in Figure ~\ref{fig:user_study}. We find that RBSR won the largest percent of votes over all six image datasets individually. RBSR decisively won the largest percentage of votes, by a margin of 10 to 55$\%$ from the second ranked algorithm, on the DIV2K HR, the RealSR-HR, the RealSR-LR, and the TV stream image datasets. The second place algorithm on these datasets alternated from RealSR, DPSR, and EDSR-Real, and RealSR respectively. We note that on the RealSR-LR dataset, for which the RealSR algorithm is tailored and trained, RBSR and EDSR-Real are the first and second place. This shows the efficacy of both the two step approach of RBSR and introducing bicubically downsampled images into the training dataset. On the DPED dataset, RBSR won by a small margin over DPSR. 

\vspace{-2mm}
\section{Conclusion}
\vspace{-1mm}
In this work, we have shown that the challenges of super resolution on realistic images can be partly alleviated by decomposing the SR pipeline into two sub-problems. First, is the conversion of real LR images to bicubic look-alike images using our novel copying mechanism and bicubic perceptual loss. Second, is the super-resolution of bicubically downsampled images. Each sub-problem addresses a different aspect of the real-world SR problem. Converting real low-resolution images to bicubic look-alike images allows us to handle and model the variety of realistic image degradations. The super-resolution of bicubically downsampled images allows for the application of state-of-the-art super-resolution models, which have achieved impressive results on images with well defined degradations. We show that our approach (RBSR) outperforms the SOTA in real-world SR both qualitatively and quantitatively using a comprehensive user study over a variety of real image datasets. 

%
\bibliographystyle{splncs04}
\bibliography{egbib}
\end{document}